# Snort® Intrusion Detection System with Intel® Software Guard Extension (Intel® SGX)

Dmitrii Kuvaiskii, Somnath Chakrabarti, Mona Vij

{Dmitrii.Kuvaiskii, Somnath.Chakrabarti, Mona.Vij}@intel.com

## Abstract

Network Function Virtualization (NFV) promises the benefits of reduced infrastructure, personnel, and management costs by outsourcing network middleboxes to the public or private cloud. Unfortunately, running network functions in the cloud entails security challenges, especially for complex stateful services.

In this paper, we describe our experiences with hardening the "king of middleboxes"– Intrusion Detection Systems (IDS) – using Intel® Software Guard Extensions (Intel® SGX) technology. Our IDS secured using Intel® SGX, called SEC-IDS, is an unmodified Snort® 3 with a DPDK network layer that achieves 10Gbps line rate. SEC-IDS guarantees *computational integrity* by running all Snort® code inside an Intel® SGX enclave. At the same time, SEC-IDS achieves *near-native performance*, with throughput close to 100% of vanilla Snort® 3, by retaining network I/O outside of the enclave. Our experiments indicate that performance is only constrained by the modest Enclave Page Cache size available on current Intel® SGX Skylake based E3 Xeon platforms. Finally, we kept the porting effort minimal by using the Graphene-SGX library OS: only 27 Lines of Code (LoC) were modified in Snort® and 178 LoC in Graphene-SGX itself.

1. Introduction

Network Function Virtualization (NFV) is a growing trend to move middleboxes such as switches, load-balancers, and firewalls from private networks into the cloud [1]. This allows companies to significantly reduce their infrastructure costs and ease resource management. The conundrum, however, is how to protect confidentiality and integrity of network functions once they are outsourced to a possibly adversarial cloud [2].

Adding protections to outsourced network functions is no easy task. To highlight its complexity, we focus on the "king of middleboxes" – Intrusion Detection Systems (IDS's) – and discuss how current research efforts fail to protect them. In particular, we dissect Snort3 – the most popular and mature open-source IDS [3].

First, IDS's are complex software systems that detect network attacks by analyzing all traffic against a set of signature-, protocol-, and anomaly-based rules. For example, Snort® decodes network packets, reassembles streams, maintains flow states, runs them through a string of protocol- and application-specific detectors, performs complex pattern matching, and signals alarms on suspicious traffic. Unfortunately, state-of-the-art crypto-scheme solutions for securing middleboxes such as BlindBox [4] and Embark [5] cannot support this rich IDS functionality.

Second, due to the aforementioned complexity of IDS's, it is undesirable to build a new system from scratch or to introduce intrusive changes in the existing codebase. Snort® has seven library dependencies, with 1 million LoC of C/C++ code; rewriting or modifying it for security purposes would be too time-consuming and error-prone. This requirement for legacy code support renders the whole middlebox-framework line of research [6–9] unsuitable for IDS's.

Third, IDS's must sustain high line rates of 10-40Gbps, e.g., a typical deployment of Snort® spawns many threads to achieve desired throughput. This requirement is at odds with software-based cryptographic solutions that inevitably degrade performance [4,5]. The better alternative is to use hardware support such as Intel® SGX [10] that provides integrity and confidentiality guarantees to applications while maintaining acceptable performance.

These three requirements dictated our approach to harden IDS's for security with the example of Snort®. To side step the complexity of Snort®, we put its entire functionality inside the Intel® SGX enclave and require (almost) no code modifications. For performance, we keep the DPDK packet acquisition layer and the packets themselves outside of the enclave and feed pointers to them to Snort® via lockless ring buffers. We call the final system SEC-IDS.

The porting effort was kept minimal by using the Graphene-SGX library OS [11]: only 27 LoC were modified in Snort® and 1,205 LoC in the DPDK layer. We also resolved minor challenges of Snort®-DPDK separation and trusted clocks along the way, leading to 178 changed LoC in Graphene-SGX.

Our evaluation shows that SEC-IDS achieves near-native speed on workloads where Snort® state fits into Intel® SGX's Enclave Page Cache (EPC), with throughput close to 100% of vanilla Snort®. Larger workload sizes can be accommodated with future versions of Intel® SGX, where the EPC size is expected to grow.

In this work, we describe the challenges of achieving near-native performance for Intel® SGX-hardened IDS (or indeed any high-performant complex application) as well as justify our design choices. In particular, we provide guidelines on cleanly separating trusted and untrusted parts of the application and avoid expensive system calls.

## 2. Background

**Intel® SGX.** The Intel® Software Guard Extensions (SGX) is an ISA extension for recent Intel® CPUs that provides confidentiality and integrity protections for sensitive parts of applications [10]. With Intel® SGX, the code and data to be protected are put inside an Intel® SGX enclave – a region of memory that is opaque to all other software including privileged OS/hypervisor. The code inside the enclave can execute almost all CPU instructions (see important exceptions below) and can access data both inside and outside of enclave. Any attempt to access enclave data from outside of the enclave fails.

At the hardware level, a handful of new x86 instructions were introduced to initialize, start, resume, and exit an enclave. When in enclave mode, the CPU disallows context switches to kernel (i.e., `syscall` and `int` instructions). Thus, when an interrupt or a system call happens, an enclave is first exited, the interrupt/syscall is handled by the kernel, and the enclave is resumed. Since an enclave exit is an expensive operation, this execution pattern leads to high performance overheads [12,13].

Another currently unsupported instruction is `rdtsc`, the low-overhead relative time source [14,15]. Unfortunately, a lot of real-world software relies on this instruction, usually through library or system calls such as `gettimeofday` and `clock_gettime`. The usual (but not secure!) workaround is to exit the enclave on `rdtsc`, execute the instruction in untrusted code, and pass the result back to the enclave [11]. Similar to syscalls, this can result in high overheads if the application uses `rdtsc` extensively.

To achieve confidentiality and integrity of enclave data, the hardware is augmented with a Memory Encryption Engine (MEE) and the Enclave Page Cache (EPC). The EPC is a designated area in physical memory inaccessible to any software other than the corresponding enclave. All data transferred from CPU to EPC is encrypted with a key associated with this particular CPU, and decrypted in another direction. Currently, EPC size is a modest 128MB with only 96MB available for user data, and enclaves exceeding this amount require expensive Intel® SGX-aware paging mechanism that securely swaps enclave pages to RAM. Because of MEE encryption and EPC paging, enclaves experience two sources of overhead: 1-10X overhead when data leaves LLC and 2-2000X when it leaves EPC [12]. Thus, it is important to keep enclave code and data to a minimum, at least on current SGX servers [7,13,16].

Other parts constituting the SGX technology, such as remote attestation, are not relevant for this paper, and we refer the reader to Intel® manuals and other sources [10].

**Graphene-SGX.** Graphene-SGX is a library OS tailored to Intel® SGX environments [11]. It allows running unmodified applications by providing a runtime to transparently exit the enclave and resume it on system calls, unsupported instructions, and interrupts. To provide additional security against, e.g., Iago attacks [17], Graphene-SGX introduces *shields* – sophisticated checks at the enclave-untrusted app interface.

Features that distinguish Graphene-SGX from similar *shielded execution* frameworks [12,14,18] include shielded dynamic loading (to support dynamically loaded libraries and run-time linking), multi-process abstractions, and file authentication. For security, a so-called *manifest* file is required with a whitelist of trusted libraries and files that can be used with Graphene-SGX. These libraries and files are also hashed, and Graphene-SGX checks that the runtime calculated hash is equal to the one specified in the manifest.

Akin to other frameworks, Graphene-SGX supports multithreaded applications (using a 1:1 threading model), exception handling, and a set of 28 Linux system calls. Graphene-SGX

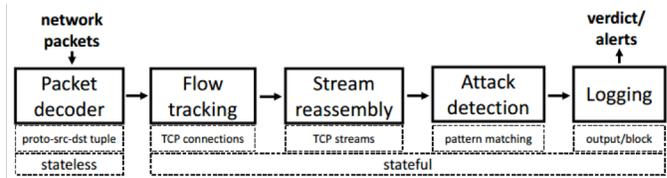

*Figure 1: Snort® IDS workflow.*

overheads are modest, with 25% lower throughput w.r.t. native for web applications.

**Intel® DPDK.** Intel® DPDK is a framework for high-speed network packet acquisition.[1] It relies on recent advances in NIC hardware and copies network packet contents directly into user space memory, completely bypassing the kernel network stack. DPDK also leverages HugePages for efficient memory management and reduced TLB pressure. This design allows achieving 10-40Gbps throughput in software middleboxes.

DPDK consists of a set of libraries that are linked into the user application. For this paper, the main libraries are Environment Abstraction Layer (EAL), `mbuf`, and `ring`. EAL abstracts away bootstrapping details of underlying NIC hardware and drivers as well as memory and thread management. The `mbuf` library defines structures and functions to store and analyze network packets (DPDK does not use Linux kernel structures). The `ring` library provides the abstraction of an RTE ring – a lockless fixed-size FIFO queue of pointers, which is primarily used for passing references to network packets.

**Snort® IDS.** Snort® is an Intrusion Detection System (IDS) that fetches packets from the network, preprocesses and analyzes them for malicious traffic [3]. In case an attack signature is detected, Snort® can either block the packet (if serving as a firewall) or generate an alert for system administrator.

Figure 1 shows the high-level overview of Snort® functionality. Each packet from the network is decoded to determine the "protocol – source IP

---

[1] http://dpdk.org/

– source port – destination IP – destination port" 5-tuple; also, the protocol fields are examined for sanity. The decoding phase is stateless.

Next, TCP packets are grouped by live connections (flows) and TCP streams are reassembled for convenience of subsequent detection modules.[2] At this point, lists of all live flows and pending segments-to-reassemble are maintained. This constitutes Snort®'s state.

After this preprocessing, the actual attack detection is invoked. Each packet payload is analyzed using two-phase pattern matching: first a simple search against a set of rules is performed. For rules that matched, the second heavyweight phase checks whether the payload contains a full attack signature, with regular expressions and flow-specific options. The *set of rules* is defined in a separate file and preprocessed at Snort® startup. We discuss Snort® rules in more detail in Appendix 1.

Finally, if an attack signature is matched, an alert is generated and output in the console or logged in a file. If Snort® is configured in inline (firewall) mode, it can also block the offending packet. If packet is innocuous, Snort® allows the packet and pushes it back to network (if in inline mode).

Snort® development started in 1998, and since then the system incorporated numerous features and became the most widely used IDS. Snort® 3 aka Snort®++ is a major rewrite of the original codebase done in C++; we use this new version in the paper. Figure 2 presents the overall architecture of Snort® 3.

Snort® is multithreaded, following the "one-thread-do-all" model. Each worker thread operates in an infinite event loop: it fetches the next packet from the network, preprocesses and analyzes it as described above, and outputs alerts. Worker threads are specifically designed to share no state for performance reasons. This design dictates that the same network flow (characterized by a 5-tuple) must be processed by the same thread; bidirectional flows require special treatment.

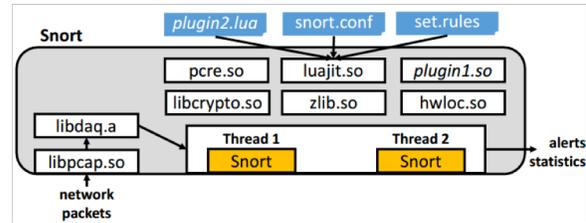

*Figure 2: Original Snort® architecture.*

Due to abundance of features, Snort® adopts a pluggable architecture and uses dynamic loading and linking extensively. It has seven library dependencies, one of which is linked statically (LibDAQ) and others dynamically. Preprocessing, analysis, and logging modules can be written as shared-library or Lua-script plugins and loaded at startup. Snort®'s *configuration file* specifies which modules and plugins to use as well as their specific settings. Due to all these features, the codebase of Snort® with all dependencies consists of 1 million LoC in C/C++ as reported by cloc.[3]

Of particular interest to us is the LibDAQ library for network packet acquisition. LibDAQ clearly separates the fetching of packets from NIC and actual Snort® processing. LibDAQ's default network library is PCAP (libpcap) – a platform-independent interface to capture packets in user-space. Due to its reliance on kernel support and interrupt-driven network I/O, it cannot handle data rates of 10-40Gbps. Alas, LibDAQ does not provide an official DPDK module as of this writing.

## 3. Threat Model and Security Properties

We assume a threat model where the only trusted entities are the CPU and the code executing in an SGX enclave, similar to other papers [11–14,16,18]. All other unprivileged and privileged software including the OS and the hypervisor is potentially malicious. Physical attacks on RAM, memory bus, or network interface are also possible. Denial-of-service and side-channel attacks (including network side channels) [19,20] are out of scope.

---

[2] Actually, Snort® keeps track of connections even for stateless protocols like UDP and ICMP, but we skip these details.

[3] https://github.com/AlDanial/cloc

| Asset | Protection |
|---|---|
| Snort® execution | Integrity |
| Snort® state (flows, streams, metadata) | Integrity and confidentiality |
| Snort® configuration and rules | Integrity and possibly confidentiality (need securely provisioned keys) |
| Network traffic | Out of scope |

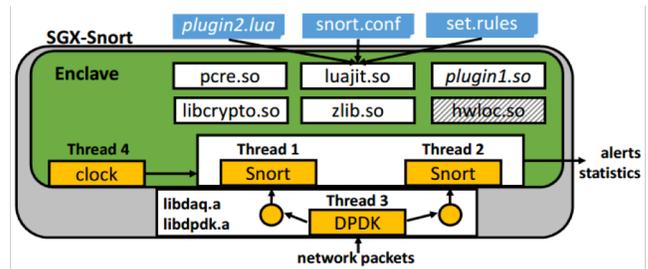

Figure 3: Architecture of SEC-IDS.

The assets we aim to protect are summarized in Table 1. In the following, we provide rationales for the proposed levels of protection.

**(No) integrity and confidentiality of network traffic.** The main motivation behind this work is to demonstrate transparency, usability and performance benefits of Intel® SGX, thus we concentrate on immediate security benefits that Intel® SGX technology and Graphene-SGX framework provide out-of-the-box. For this reason, we do not target *confidentiality of network traffic*, where all traffic passing through the middlebox is encrypted but still correctly processed. This property is not trivial to achieve for an IDS like Snort® that expects packets in the clear. In theory, packets could arrive encrypted and then be decrypted by Snort® before processing, but such a setup would require provisioning all session keys to Snort®, which is infeasible.

We also do not aim to protect *integrity of network packets*. The attacker can drop or rearrange packets as well as inject malicious packets or modify parts of the true ones (since in our scenario traffic is in plain text). A low-cost weak-security solution would be to install a simple statistics-collecting middlebox at user's premises and compare against statistics collected by enclavised Snort®. However, there is no general way to protect integrity of packets: a powerful network attacker can simply modify incoming or outgoing packets before or after Snort® inspection.[4]

**Integrity of execution and configuration.** We aim to protect *integrity of execution* (aka *computational integrity*) of Snort®: the user is assured that Snort® runs correctly with correct configuration and set of rules. This is useful to verify functionality and performance of Snort® for audit purposes. In addition, Intel® SGX protects the confidentiality of the internal state of Snort® including metadata, flows, and streams. Note that it is trivial to add support for *confidentiality of rules and configuration* by providing encrypted rules/configuration files and decrypting them inside an enclave with a securely provisioned key.

**System call interface.** The OS can potentially launch attacks on the enclave using the system call interface, i.e., Iago attacks [17]. The Graphene-SGX framework already provides shields for 28 most commonly used system calls, but, as we will see, Snort® relies on additional syscalls for correct functioning. Thus, it is necessary to carefully examine the semantics of these syscalls. We revisit this requirement in Section 4.

---

[4] For a network attack to succeed, the attacker needs access to a physical device or powerful OS privileges. This can be easily detected, unlike more subtle attacks on Snort® that we try to prevent in this work.

## 4. Architecture

Our architectural goal is to run Snort® inside an Intel® SGX enclave with near-native performance and minimal changes to the code base. To accomplish this, we chose the path of iterative refinement, first putting all Snort® code inside the enclave and then adding features one by one. After five iterations, we came up with the final architecture presented on Figure 3. In the following, we discuss encountered challenges and our solutions.

**Challenge 1: Snort® and its dependencies inside enclave.** As a first step, we needed to port all Snort® functionality, including library dependencies and loadable plugins, inside an enclave with minimum effort.

A naïve solution would be to use the Intel® SGX Software Development Kit (SDK).[5] The SDK provides building blocks to create, run, and attest enclaves, seal data to be stored on disk, and tools to develop ECALL/OCALL enclave interfaces. Even though the SDK can be used to port complex existing applications, the manual effort to define all required interfaces and to modify the original code base is significant (e.g., 11,078 LoC of changes in SGX-Tor) [21]. Manually porting huge code bases is known to be error-prone, leading to new vulnerabilities and negating the promise of Intel® SGX-enabled security.

The appropriate solution is to use a shielded execution framework that hides the complexity of adjusting applications to enclaves. There is a number of frameworks to choose from: Haven [14], Graphene-SGX [11], SCONE [12], and Panoply [18]. Haven is an early closed-sourced Windows-based Library OS system with a large 210MB memory footprint. Graphene-SGX is a Library OS for Linux applications with a much smaller TCB. SCONE and Panoply decrease TCB even further but do not support dynamic linking and loading required by Snort®. In fact, SCONE's features of user-level threading and asynchronous syscalls are ineffective in the Snort® + DPDK case: Snort® invokes *no* system calls during normal execution (except `clock_gettime` that we discuss next). Lastly, Panoply prioritizes minimal TCB over performance and thus exhibits higher overheads than Graphene-SGX.

Ultimately, we settled on Graphene-SGX since it perfectly aligns with Snort®'s requirements: it supports dynamic linking and loading, OS signals, 1:1 multithreading, and file authentication. On top of this, Graphene-SGX provides high performance and strong security via syscalls' shields.

Even with Graphene-SGX, porting Snort® required some code and build-system modifications. We encountered three minor issues:

*(i)* The hwloc library – to set affinity of Snort® worker threads for better performance – issues `sched_setaffinity` and `sched_getaffinity` system calls. The kernel is supposed to set and get a CPU affinity mask, but there is no way to prove correctness of these actions from within the enclave. Thus, issuing these syscalls to a malicious OS is at best futile and at worst can trigger obscure application bugs (e.g., Snort® assumed that `sched_getaffinity` never fails and went into infinite loop when we tried to stub the syscall with a dummy value). In the end, we decided to remove the hwloc dependency from Snort® altogether, patching 27 LoC in a single file.[6]

---

[5] https://software.intel.com/en-us/sgx-sdk

[6] This was the only change we made in Snort® itself. We would even argue that this was beneficial for SEC-IDS: by removing hwloc, we instantly decreased TCB by 40,000 LoC.

*(ii)* The luajit library contains an embedded Lua interpreter to parse Snort®'s configuration file and Lua-based plugins. The library uses a rare `MAP_32BIT` flag when allocating memory via `mmap`. This flag instructs the OS to allocate memory in first 2GB of process address space, for outdated performance reasons. Graphene-SGX did not support this flag, thus we introduced a one-line patch to Graphene to ignore it inside the enclave. Note that this change does not affect security since enclaves use their own memory management anyway.

*(iii)* The libpcap library is used for user-level packet capture in the PCAP format. It operates on so-called "raw sockets" to fetch IP packets. The only types of sockets Graphene-SGX supports are TCP/UDP, and adding shielding support for a too-powerful primitive like raw sockets would go against the LibOS philosophy. We removed libpcap as a dependency since SEC-IDS already uses the DPDK library residing outside of the enclave for packet acquisition.

These were the only changes required to run Snort® and its dependencies inside an Intel® SGX enclave.

**Challenge 2: DPDK outside enclave.** Our next challenge was to provide an interface to allow communication between Snort® threads inside the enclave and DPDK threads outside of it.

We first need to explain our decision to leave DPDK (and LibDAQ for that matter) outside of the enclave. After all, it is possible to have DPDK code and state inside the enclave. However, we see three problems with this design. First, it would still require adding low-level networking support to Graphene-SGX, similar to libpcap. Second, putting all DPDK code – 113,000 LoC – inside the enclave unnecessarily bloats TCB. Third, to store fetched packets, DPDK uses HugePages not supported by Intel® SGX, and there is little sense to protect DPDK code and metadata while the packets themselves are left in untrusted memory.

As of this writing, there is no official support for DPDK in Snort® 3. Thus, we used a fork of the LibDAQ library with a DPDK module from Napatech.[7] The library is feature-rich and allows to correctly load-balance network flows across multiple Snort® threads, using the Receive Side Scaling (RSS) network driver technology [22].

The original design of this DPDK module does not suit our needs: it combines DPDK packet fetching and Snort® processing in one thread ("one-thread-do-all" model). In other words, each worker thread runs an infinite loop with three steps: (1) receive packets from network in a burst, (2) run Snort® analysis on each received packet, and (3) transmit each allowed packet back to network, if Snort® is in inline mode.

Since we want to run DPDK code outside of the enclave, we break this functionality into two parts (compare Figures 2 and 3). We designate M "DPDK threads" that perform steps 1 and 3 as above. However, instead of directly calling into Snort®, each DPDK thread puts pointers to fetched packets in a RX ring and gets pointers to allowed packets from a TX ring, if in inline mode. A second set of N "Snort® threads" perform step 2, i.e., actual Snort® analysis. Each Snort® thread gets the next available pointer from a RX ring, analyzes the packet pointed to, and puts the pointer to this packet in a TX ring if the packet is allowed.

RX and TX rings are implemented as RTE lockless rings and serve as a main communication channel between the enclavized Snort® and outside DPDK. They are created at startup and allocated in unprotected HugePages memory. Note that these rings contain only pointers to packets (more precisely, to their `mbuf` DPDK

---

[7] https://github.com/napatech/daq_dpdk_multiqueue

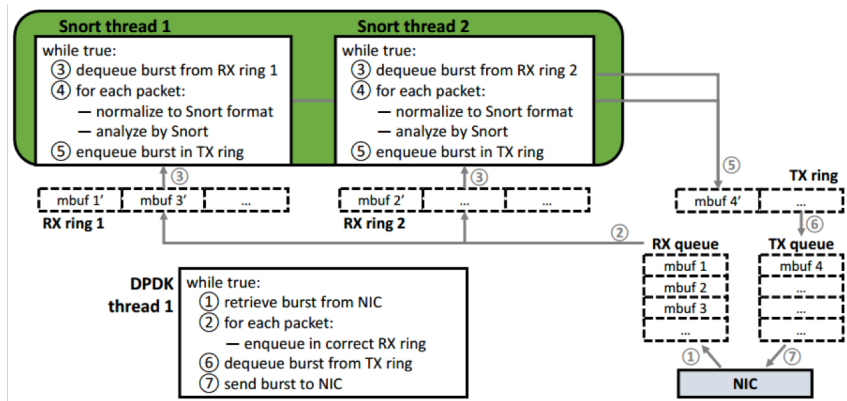

*Figure 4: Interface between Snort® and DPDK threads via RX/TX rings.*

representations), allowing zero-copy packet analysis.

There is one caveat with RX rings. Each Snort® thread operates on its own subset of network bidirectional flows. Thus, we cannot use a single shared RX ring: all Snort® threads would read from this ring in no particular order and network flows will become jumbled. To alleviate this, we create as many RX rings as there are Snort® threads. For every fetched packet, a DPDK thread looks into the last six bits of the RSS hash and maps them to the corresponding RX ring. Since all packets in the same network flow produce the same RSS hash, they end up in only one Snort® thread.

Pushing packets back to the network requires only a single shared TX ring. This is why we use a Multiple Producer Multiple Consumer (MPMC) TX ring, but allow Multiple Producer Single Consumer (MPSC) RX rings.

Figure 4 illustrates the interface between Snort® and DPDK threads via RX and TX rings. Note that the rings contain pointers to `mbuf` objects, denoted with an apostrophe. Also note that Snort® sends packets via TX ring and TX queue only in inline mode.

Using separate DPDK and Snort® threads entails two additional benefits. First, since each of these threads runs on a separate CPU core, the L1 cache is less stressed than in the "one-thread-do-all" design. Second, our experiments indicate that a single DPDK thread is enough to sustain 10Gbps network load.

| OCALL name | Description |
|---|---|
| dpdk_initialize | Process command-line arguments, allocate memory for packets and RX/TX rings, initialize EAL |
| dpdk_start_device | Set up RX and TX queues for Ethernet device, set RSS hash for bidirectional flows, start Ethernet device |
| dpdk_acquire | Start infinite loop of DPDK threads: retrieve burst of packets from network, pass to RX ring, read from TX ring and push allowed packets back to network |
| dpdk_stop | Close Ethernet device |
| dpdk_shutdown | Close Ethernet device and free all allocated memory |

*Table 2: OCALL interface between Snort® and DPDK.*

**Challenge 3: Interfaces between Snort® and DPDK.** There are two interfaces between the enclavized Snort® and untrusted DPDK: (1) RX/TX rings to pass pointers to network packets, and (2) OCALLs to initialize and shutdown LibDAQ and DPDK.

The interface of RX/TX rings was described in the previous section. It is the only interface during run-time, with asynchronous "exitless" communication. This interface is what makes SEC-IDS perform on par with vanilla Snort®.

For initialization and finalization of DPDK, SEC-IDS requires another interface with function call invocations. In particular, we introduce five

OCALLs from enclave to untrusted code (see Table 2).

We manually add these OCALLs to Graphene-SGX and patch the DPDK module in LibDAQ to call them. The typical workflow is thus as follows (with the example of `dpdk_initialize`): (1) Snort® starts inside the enclave, (2) it calls LibDAQ initialization routine, (3) LibDAQ recognizes that it is inside the enclave and calls `ocall_dpdk_initialize`, (4) this OCALL exits the enclave, performs untrusted `dpdk_initialize`, and resumes enclave execution.

Note that we link LibDAQ both inside and outside of the enclave. The enclavized LibDAQ is compiled to serve as a minimal redirection layer: it only invokes OCALLs. The outside LibDAQ performs actual DPDK initialization/finalization.

The five OCALLs are invoked only on SEC-IDS startup and teardown and do not affect performance.

**Challenge 4: Trusted clock.** After we separated enclavized Snort® and untrusted DPDK threads, we made sure no system calls were issued by the application. To our surprise, the overhead of SEC-IDS over vanilla Snort® still was 1,000%!

Upon further examination, we noticed an unusually high number of enclave exists and resumes. Graphene-SGX exits the enclave only on system calls, but we observed none via strace, so what was happening?

The root cause turned out to be the `clock_gettime` system call. In usual environments, `clock_gettime` is virtualized with vDSO library, i.e., instead of context-switching to the kernel, this system call is resolved in user space via the `rdtsc` instruction. In Graphene-SGX, since `rdtsc` is disallowed inside enclaves, `clock_gettime` is treated as a syscall: upon its invocation, Graphene-SGX exits the enclave, executes the virtualized version outside, and resumes secure execution. Thus, we observe

| Software | Total LoC | Changed LoC |
|---|---|---|
| Snort® | | |
| Snort® binary | 228,633 | 27 (0.02%) |
| LibDAQ | 40,523 | 1,205 (3%) |
| hwloc | 40,397 | N/A * |
| other libs | 701,234 | 0 (0%) |
| *total* | *1,010,787* | 1,232 (0.12%) |
| Graphene-SGX | | |
| Graphene binary | 1,233,484 | 0 (0%) |
| new OCALLs | 157 | 157 (100%) |
| clock thread | 21 | 21 (100%) |
| *total* | *1,233,662* | 178 (0.01%) |

*Table 3: Lines of code in all used software.*
*\* hwloc was removed from dependencies.*

many enclave exits (that degrade performance) but no actual system calls.

Snort® heavily relies on `clock_gettime`, invoking it *at least twice* for each packet: it is used to measure timeouts, TCP/UDP flow expirations, packet latencies, and passage of time for statistics. Thus, we could not eliminate this syscall and had to find a performant workaround. Note that if we would execute `rdtsc` outside and pass the resulting clock back to the enclave, we would be susceptible to Iago attacks (the hacker could easily subvert Snort® execution by providing bogus time values).

In the end, we settled for a "trusted-clock" thread, technique described in the "Malware Guard Extension" paper [15]. We introduce a helper "clock" thread in Graphene-SGX that runs inside the enclave and infinitely increments a global variable. Additionally, we stubbed `clock_gettime` to read the value of this variable, adjust it to our CPU's speed, and return time in microseconds. This simple technique provides a "good enough" relative time source for our purposes. Details can be found in Appendix 2.

## 5. Implementation Details

We used the following versions of software: Intel® DPDK 17.05.1, Intel® SGX Driver v1.9 (commit 3abcf82), Graphene-SGX commit 4d8eacd, Napatech's fork of LibDAQ v2.2.1 (commit 7c40e02), Snort3 tag BUILD_239, LuaJIT v5.1, PCRE v8.41, zlib v1.2.11, OpenSSL v1.1.0.

All changes in Snort® and Graphene-SGX are summarized in Table 3. We expand on the changes below.

In Snort® codebase, we modified 27 LoC in one file (`thread_config.cc`) to disable the dependency on hwloc.

In Napatech's LibDAQ, we modified 1,205 LoC in one file (`daq_dpdk.c`) to introduce the ECALL interface and add more functionality like RSS-based mapping to Snort® threads. For experiments, we also added code to manually pin Snort® threads to physical cores.

In Graphene-SGX, we made several changes: (1) one-line change to support `MAP_32BIT`, (2) adapted the build system to link against untrusted LibDAQ and DPDK, (3) added 157 LoC for five OCALLs, and (4) added 21 LoC for trusted-clock thread.

Graphene-SGX requires all its dependencies to be built with `-fPIC` flag (as position-independent libraries). We added this flag to DPDK and LibDAQ builds and observed no drop in performance. For enclavized Snort® build, we link it against the "dummy" LibDAQ that only redirects function calls to the outside LibDAQ via OCALLs.

We identified and fixed a multithreading bug in Graphene-SGX that led to premature exhaustion of enclave thread slots. We also added an `exit_group` syscall to terminate all process threads – this was needed to kill our trusted-clock thread. Finally, we observed an infrequent data race somewhere in the memory-allocation logic of Graphene, but could not pinpoint its root cause.

## 6. Evaluation

In our evaluation, we aim to answer the following questions:

- What is the overhead of SEC-IDS with respect to vanilla Snort® in terms of achievable throughput?
- What is the effect of the packet size and the number of flows in workload?
- What is the effect of the number of rules, enabled functionality, and logging in Snort® configuration?
- What is the scalability of SEC-IDS?

**System platform.** We use two servers connected via a 10Gbps NIC. Each server has an Intel® Xeon® CPU E3-1270 v5 @ 3.60GHz with one socket, 4 physical cores and hyper-threading disabled, 64GB of DDR4 2133Mhz RAM, 8MB L3 cache, 256KB per-core L2 cache, and 32KB instruction and data per-core L1 caches. The NIC is X710 for 10GbE SFP+ 1572, and we use vfio-pci DPDK driver. We use Ubuntu 16.04 with kernel v4.4.0. We assign 16 1GB-sized HugePages (totaling 16GB of RAM) and pin DPDK and Snort® threads to dedicated cores. For workload generation, we use Intel® PktGen v3.4.0.

**Methodology.** We ran experiments on our SEC-IDS as well as vanilla Snort®. By vanilla Snort® we mean the Snort®+DPDK version with LibDAQ and DPDK code as described above but without SGX enclaves (and thus without Graphene-SGX). We use default configuration file of Snort® and the set of rules `community.rules` from the official web-site. We also use the default build system of Snort®.

We use two workload scenarios: synthesized and real-pcap ones. Synthesized workloads are TCP/IP packets with random payloads generated by PktGen. We vary packet sizes from 64B to 1024B, number of simultaneous network flows (TCP connections) from 256 to 32,000, and number of Snort® rules used for pattern matching from 0 to 3462 (all rules in `community.rules`). Note that the number of flows in a real network is much higher, typically millions of simultaneous flows; however, PktGen segfaulted on more than 32,000 flows.

We also use three real PCAP workloads, downloaded from Tcpreplay.[8] The characteristics of these workloads are as follows: (1) `test.pcap` contains 141 packets with 37 flows and average packet size 445B, (2) `smallFlows.pcap` contains 14,261 packets with 1,209 flows and average packet size 646B, and (3) `bigFlows.pcap` contains 791,615 packets with 40,686 flows and average packet size 449B. These three workloads stress Snort® rules and output real alerts.

Each experiment is run for 2 minutes: we first initialize PktGen and start sending workload packets (PktGen repeats packets if it reaches the end of a PCAP file), then start Snort®, wait for 2 minutes, send a SIGINT signal to Snort® to gracefully terminate, log all Snort® output, and stop PktGen. Snort® output contains statistics on packets received, analyzed, and allowed, as well as throughput numbers. We use these statistics to draw our plots. Appendix 3 has details on how exactly we run Snort® and PktGen.

Each experiment was run three times, and we use a standard mean across three runs. The standard error is less than 1% in all experiments.

**Results with synthesized workloads.** Figures 5-7 show Snort® throughput with varied packet sizes, number of flows, and number of rules.

One immediate thing to note: in many cases, SEC-IDS performs slightly better than vanilla Snort®. The reason is our trusted-clock thread in the Intel® SGX version; the vanilla version uses regular `rdtsc`. The trusted-clock mechanism induces a lower latency per `rdtsc`, thus SEC-IDS executes `clock_gettime` slightly faster than vanilla.

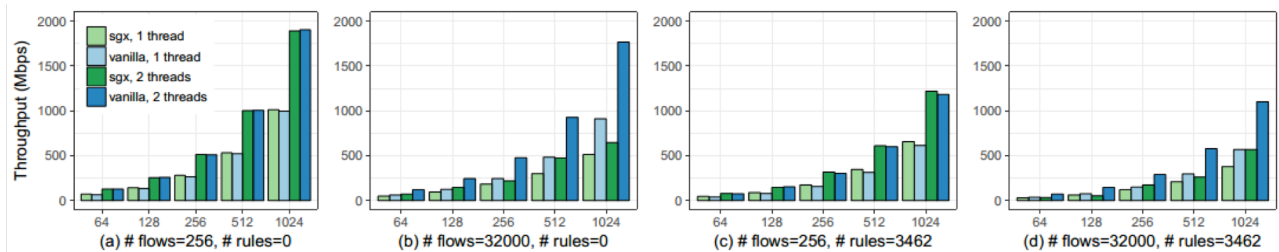

Figure 5: Throughput of SEC-IDS and vanilla SGX with increasing packet size.

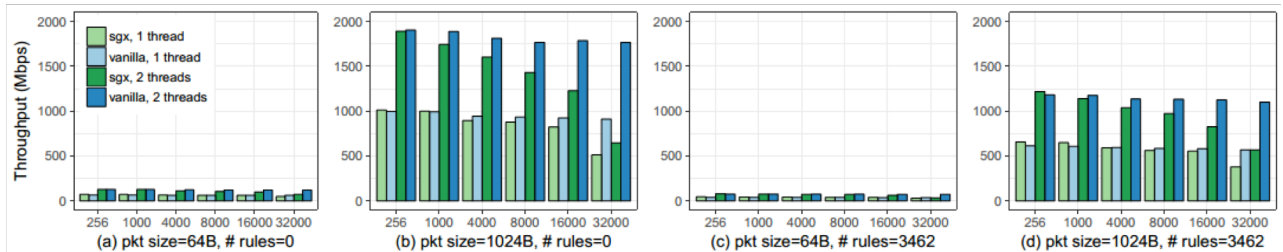

Figure 6: Throughput of SEC-IDS and vanilla SGX with increasing number of TCP flows.

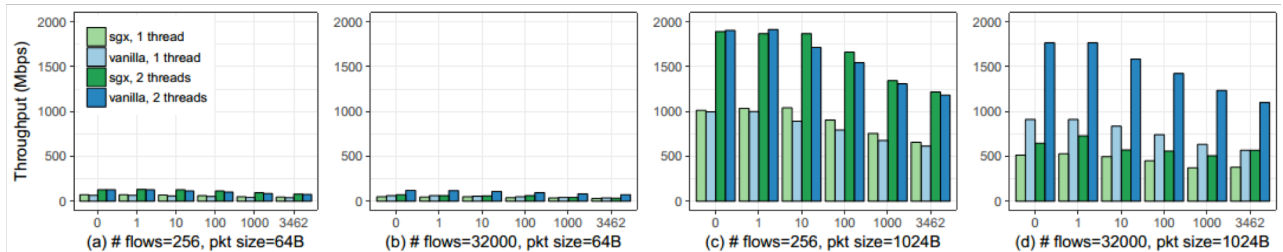

Figure 7: Throughput of SEC-IDS and vanilla SGX with increasing number of rules (max 3462).

We chose to show four extreme points for each varied parameter. For example, Figure 5 has packet size on X-axis and four combinations of # flows and # rules. On Figure 5a, we show the "performant" configuration with a tiny number of flows (256) and no rules at all (i.e., no pattern matching). This configuration clearly has the best performance for both Snort® and SEC-IDS: the state that Snort® keeps is very small because the number of flows is very small. The processing per packet is very fast because there is no pattern one thread and 812MB for two threads) because there are 32K flows and processing per packet is slow because it requires matching of all 3462 rules. Figures 5b and 5c show intermediate points: with many flows but no rules and with tiny number of flows but all rules. Figures 6 and 7 follow the same pattern.

It is clear that increasing *packet sizes* also increase throughput of both vanilla and SGX Snort®. As the packet size increases, the ratio of

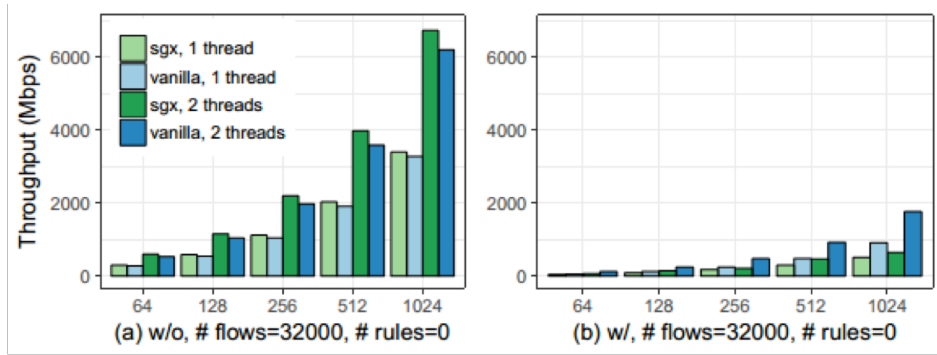

*Figure 8: Throughput of SEC-IDS and vanilla SGX without and with configuration file (increasing packet sizes).*

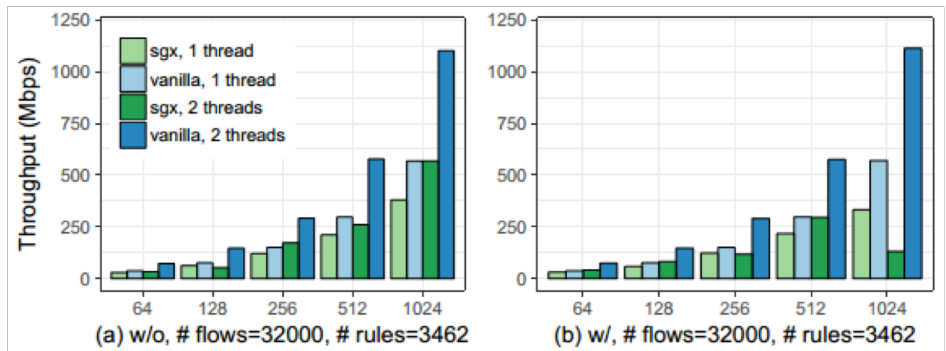

*Figure 9: Throughput of SEC-IDS and vanilla SGX without and with outputting alerts (increasing packet sizes).*

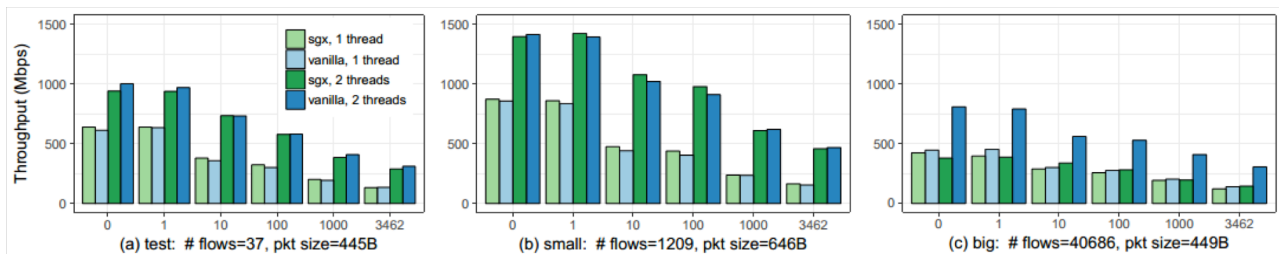

*Figure 10: Throughput of SEC-IDS and vanilla SGX with real-world PCAP traces (increasing number of rules).*

matching. On Figure 5d, we see another extreme: Snort®'s state is large (up to 620MB for

Snort® processing to "bytes received" decreases. In other words, the overhead is determined by the number of packets and not by the number of bytes per packet.

Figure 6 indicates that increasing *number of flows* has no effect on vanilla Snort® but hampers performance of SEC-IDS. It hints at the Intel® SGX limitation: storing a flow in Snort® requires 2-4KB of memory, and while vanilla Snort® is not limited by RAM, SEC-IDS quickly exhausts the EPC size and requires expensive EPC paging. This is proved by the CPU utilization of the Intel® SGX swapping daemon (ksgxswapd): with 32,000 flows it peaks to 12% for two threads and 24% for three threads (while in other cases it is zero). We also notice that SEC-IDS performs well with one thread but suffocates with two threads, clearly seen on Figures 6b and 6d. This is because two Snort® threads do not share any state and thus require (partial) duplication of memory.

Figure 7 shows the impact of the *number of rules*. We take first N rules from the `community.rules` file for this experiment, with the maximum of 3,462 rules. Here we observe that throughput drops slightly in both vanilla and SGX Snort®, but there is no difference between vanilla and Intel® SGX versions (clearly seen on Figure 7c). This is understandable: the whole set of rules occupies only 28MB and does not stress memory.

Finally, we can compare *1-thread and 2-thread*s versions on Figures 5-7. Vanilla Snort® shows perfectly scalability, doubling its throughput when moving from a single thread to two threads. SEC-IDS has a more complex behavior. With a small number of flows, SEC-IDS also scales perfectly. However, as the number of flows increases, the scalability of SEC-IDS declines; this is seen especially in Figures 6b and 6d. We highlighted the reasons for this above.

As an additional experiment, we looked at the effect of the *configuration file*. Figure 8 shows two plots: without a configuration file on the left and with it on the right. Running Snort® without configuration means that it does not perform any preprocessing or analysis of packets: it simply fetches packets from the RX ring and immediately allows them (aka useless mode). Clearly, without any processing, Snort® does not keep any state and has almost-zero latency for each packet. As soon as we add the configuration file, Snort® runs at full capacity, and performance drops for both vanilla and Intel® SGX versions. Note that Intel® SGX version performs poorly in Figure 8b because we stress it with 32,000 flows.

Figure 9 shows the effect of *enabling alerts*, i.e., showing them in the console. There is no performance degradation for both vanilla and SGX Snort® except one case: SEC-IDS is 4X worse with alerts on 1024B packets and 2 threads (right-most bar). This is again due to the EPC paging: at this point, `ksgxswapd` CPU utilization jumps from 25% to 30%.

Finally, we experimented with three real-world PCAP traces: test (very small, only 37 flows), small (1,209 flows), and big (40,686 flows). Figure 10 shows the results, with the increasing number of rules. Similar to observations on synthesized workloads, SEC-IDS performance overhead is a function of the number of flows: with a `big` PCAP and two threads, throughput is roughly halved.

In general, we conclude that the only source of performance degradation of SEC-IDS is the limited EPC size. This is clear when we stress the system with 32,000 flows and a 2-threads configuration. In low-memory-footprint cases, SEC-IDS performs very similar to vanilla Snort®. Additional plots can be found in Appendix 4.

## 7. Related Work
**Intel® SGX-enabled middle boxes.** The benefits of Intel® SGX for network middleboxes were explored in previous works, though never at a scale of IDS's [7—9,23]. For example, LightBox [7] introduces a new framework for Intel® SGX-enabled network functions, with secure configuration and data channels and a smart state cache to reduce the negative impact of EPC paging. In a similar vein, TrustedClick [8] and Slick [9] add secure communication channels with remote attestation to the existing Click architecture. None of these projects aims to support unmodified complex IDS's such as Snort®.

S-NFV [23] manually separates only the state-handling code and data (kept in the enclave) from the rest of application. As one case study, S-NFV modifies Snort® to keep tag state in the enclave: the results indicate huge developer effort and high overheads for a small piece of functionality. It is unclear how this approach scales to flow states and stream reassembly, as well as what security implications result from leaving most of Snort® code and data outside of enclave.

**IDS's and Snort®.** Research on IDS's and Snort® in particular gained a lot of attention from the security community. To the best of our knowledge, all these efforts concerned increasing performance and *not* security of Snort®. For example, Kargus reengineers Snort® using (1) batch processing at all stages and (2) CPU-GPU parallel execution for pattern matching [24]. Similarly, GASPP rewrites Snort® to run completely on GPUs for performance benefit [25]. We are not aware of any attempts to harden complete Snort® for security.

**Intel® SGX-enabled network applications.** The work closest to ours in spirit is the recent SGX-Tor [21]. Similar to our work, SGX-Tor moves security-critical functionality of the Tor anonymity network application inside an Intel® SGX enclave. Unlike our work, SGX-Tor modifies a significant 3.4% (or 11,078 LoC) of the original code and uses the Intel® SGX SDK for manual separation. Finally, SGX-Tor proves that a smart partitioning into enclave and untrusted parts leads to low performance overheads of 4-12%.

## 8. Lessons Learned

**Support for manual separation in shielded execution frameworks.** Our Snort® case study highlights an interesting hybrid of *unmodified execution* and *manual separation*. On the one hand, it would be tedious and error-prone to reorganize a mature application like Snort® to use Intel® SGX enclaves, thus a shielded execution framework is a natural choice. On the other hand, the developer might want to tweak the application to achieve better performance by moving some parts outside of the enclave.

Based on our experience with manual Snort®-DPDK separation in Graphene-SGX, we believe that a generic extension to add application-specific OCALLs would be beneficial and easy to implement. Even in our case, with Graphene-SGX not designed for manual OCALLs, we were able to achieve clear separation in only 157 LoC. Expanding Graphene-SGX and probably combining it with the flexibility of the Intel® SGX SDK would be a step in the right direction.

**Manual separation at the library interface.** Our experience with Snort® and LibDAQ proves an apparent intuition: separation between enclave and untrusted code/data is easy at the library interface. Libraries are designed to be self-contained and to expose only a minimal set of functions. This provides an obvious choice for OCALLs. Additionally, libraries usually expose their state in a well-defined set of struct-objects; this shared state is allocated by the library outside of the enclave and the pointer to it should be passed to the enclave.

**Performance limitations of current Intel® SGX servers are temporary.** Many current limitations – small EPC size, unavailability of the `rdtsc` instruction – will be addressed in future implementations of Intel® SGX. Our work on SEC-IDS indicates that there will be no tangible performance loss once the EPC size is sufficiently large.

# Bibliography

[1] Justine Sherry, Shaddi Hasan, Colin Scott, Arvind Krishnamurthy, Sylvia Ratnasamy, and Vyas Sekar. Making middleboxes someone else's problem: network processing as a cloud service. SIGCOMM'2012

[2] Luca Melis, Hassan Jameel Asghar, Emiliano De Cristofaro, Mohamed Ali Kaafar. Private Processing of Outsourced Network Functions: Feasibility and Constructions. arXiv:1601.06454, 2016

[3] Martin Roesch. Snort® - Lightweight Intrusion Detection for Networks. LISA'1999

[4] Justine Sherry, Chang Lan, Raluca Ada Popa, and Sylvia Ratnasamy. BlindBox: Deep Packet Inspection over Encrypted Traffic. SIGCOMM'2015

[5] Chang Lan, Justine Sherry, Raluca Ada Popa, Sylvia Ratnasamy, and Zhi Liu. Embark: securely outsourcing middleboxes to the cloud. NSDI'2016

[6] Aurojit Panda, Sangjin Han, Keon Jang, Melvin Walls, Sylvia Ratnasamy, and Scott Shenker. NetBricks: taking the V out of NFV. OSDI'2016

[7] Huayi Duan, Xingliang Yuan, Cong Wang. LightBox: SGX-assisted Secure Network Functions at Near-native Speed. arXiv:1706.06261, 2017

[8] Michael Coughlin, Eric Keller, and Eric Wustrow. Trusted Click: Overcoming Security issues of NFV in the Cloud. SDN-NFVSec'2017

[9] Bohdan Trach, Alfred Krohmer, Sergei Arnautov, Franz Gregor, Pramod Bhatotia, Christof Fetzer. Slick: Secure Middleboxes using Shielded Execution. arXiv:1709.04226, 2017

[10] Costan, Victor, and Srinivas Devadas. Intel® SGX Explained. IACR Cryptology ePrint Archive 2016

[11] Chia-Che Tsai, Mona Vij, and Donald Porter. Graphene-SGX: A Practical Library OS for Unmodified Applications on SGX. USENIX ATC'2017

[12] Sergei Arnautov, Bohdan Trach, Franz Gregor, Thomas Knauth, Andre Martin, Christian Priebe, Joshua Lind, Divya Muthukumaran, Dan O'Keeffe, Mark L. Stillwell, David Goltzsche, David Eyers, Rüdiger Kapitza, Peter Pietzuch, and Christof Fetzer. SCONE: secure Linux containers with Intel® SGX. OSDI'2016

[13] Meni Orenbach, Pavel Lifshits, Marina Minkin, and Mark Silberstein. Eleos: ExitLess OS Services for SGX Enclaves. EuroSys'2017

[14] Andrew Baumann, Marcus Peinado, and Galen Hunt. Shielding applications from an untrusted cloud with Haven. OSDI'2014

[15] Michael Schwarz, Samuel Weiser, Daniel Gruss, Clémentine Maurice, Stefan Mangard. Malware Guard Extension: Using SGX to Conceal Cache Attacks. arXiv:1702.08719

[16] Dmitrii Kuvaiskii, Oleksii Oleksenko, Sergei Arnautov, Bohdan Trach, Pramod Bhatotia, Pascal Felber, and Christof Fetzer. SGXBOUNDS: Memory Safety for Shielded Execution. EuroSys'2017

[17] Stephen Checkoway and Hovav Shacham. 2013. Iago attacks: why the system call API is a bad untrusted RPC interface. ASPLOS'2013

[18] Shweta Shinde, Dat Le Tien, Shruti Tople, Prateek Saxena. Panoply: Low-TCB Linux Applications with SGX Enclaves. NDSS'2017

[19] Shweta Shinde, Zheng Leong Chua, Viswesh Narayanan, and Prateek Saxena. Preventing Page Faults from Telling Your Secrets. ASIA CCS '2016

[20] Marcus Haehnel, Weidong Cui, and Marcus Peinado. High-Resolution Side Channels for Untrusted Operating Systems. USENIX ATC'2017

[21] Seongmin Kim, Juhyeng Han, Jaehyeong Ha, Taesoo Kim, and Dongsu Han. Enhancing Security and Privacy of Tor's Ecosystem by Using Trusted Execution Environments. NSDI'2017

[22] Shinae Woo, Kyoungsoo Park. Scalable TCP Session Monitoring with Symmetric Receive-side Scaling. Technical Report. 2012

[23] Ming-Wei Shih, Mohan Kumar, Taesoo Kim, and Ada Gavrilovska. S-NFV: Securing NFV states by using SGX. SDN-NFV Security'2016

[24] Muhammad Asim Jamshed, Jihyung Lee, Sangwoo Moon, Insu Yun, Deokjin Kim, Sungryoul Lee, Yung Yi, and KyoungSoo Park. Kargus: a highly-scalable software-based intrusion detection system. CCS'2012

[25] Giorgos Vasiliadis, Lazaros Koromilas, Michalis Polychronakis, and Sotiris Ioannidis. GASPP: a GPU-accelerated stateful packet processing framework. USENIX ATC'2014

## Appendix 1: Snort® Rules

Snort® rules are used to detect anomalies in the network, malicious packets, and hacker attacks. Most Snort® rules are written as regular expressions with additional protocol/flow properties attached. Rules that are more complicated are written in C++ or Lua and distributed with Snort® or as separate plugins.

Sets of rules are compiled in a single file loaded at Snort® startup. These files can be freely downloaded from an official Snort® web-site (so-called "community" rules). There are also paid rule subscriptions that are updated more frequently and contain extra rules. In this work, we use the "community.rules" file.

As of date of this writing, the community set contains 3,462 rules. A typical rule is a one-liner that looks as follows:

```
alert tcp $HOME_NET [21, 25, 443, 465, 636, 992, 993, 995, 2484] -> $EXTERNAL_NET any (msg: "OpenSSL SSLv3 large heartbeat response - possible ssl heartbleed attempt"; flow: to_client, established, only_stream;  content: "|18 03 00|", depth 3;   byte_test: 2,>,128,0,relative; metadata: policy balanced-ips drop, policy security-ips drop, ruleset community;  service:  ssl;  reference:  cve,2014-0160; classtype: attempted-recon; sid: 30514; rev: 9; )
```

The above rule detects a Heartbleed attack and generates an alert with a message `msg`. Snort® will examine all TCP packets with source ports 21, 25, 443, etc. (typical ports that use SSL/TLS) that are sent to client over an established connection, with first 3 bytes containing `18 03 00` (message type) and the next 2 being greater than "128" (i.e., payload length is greater than 128 bytes – indication of Heartbleed). Additionally, the rule instructs to drop the packet if in Intrusion Prevention (inline) mode. The rest fields indicate CVE of vulnerability, rule set, and rule SID and version.

## Appendix 2: Trusted-clock Thread

We implemented the trusted-clock thread using the following assembly code. We changed Graphene-SGX to start an additional enclave thread that infinitely executes `clock_thread_main`. The `trusted_clock` global variable serves as a time source.

```
volatile long unsigned trusted_clock;
int clock_thread_main(void* unused) {
   trusted_clock = 0;
   asm volatile (
       "mov %0, %%rcx\n\t"
       "mov (%%rcx), %%rax\n\t"
       "1: inc %%rax\n\t"
       "   mov %%rax, (%%rcx)\n\t"
       "   jmp 1b"
       : /* no output operands */
       : "r"(&trusted_clock)
       : "%rax", "%rcx", "cc"
       );
   return 0;
}
```

We patch the `ocall_gettime` system call to access `trusted_clock` and adjust it to return time in microseconds.

```
int ocall_gettime (unsigned long * microsec) {
   #define CPUFREQ 3785.0
   extern volatile long unsigned trusted_clock;
   *microsec = (long unsigned) (trusted_clock/CPUFREQ);
   return 0;
}
```

For our prototype, we skipped the issues around integer overflow and clock drift. We also hardcode the coefficient `CPUFREQ` for adjustment to reflect our particular machine. Note a subtle security implication: the malicious OS can preempt the clock thread at will and artificially slow down the passage of time. However, the OS cannot revert passage of time, since the counter only *monotonically increases*.

We consider our clock-thread feature a temporary workaround, since ultimately the `rdtsc` issue should be fixed in Intel® SGX hardware.

## Appendix 3: Commands for Experiments

First, we modified the GRUB2 boot loader's settings for Linux:

```
GRUB_CMDLINE_LINUX = "default_hugepagesz=1GB hugepagesz=1G hugepages=16 iommu=pt intel_iommu=on intel_idle.max_cstate=0 intel_pstate=disable isolcpus=2-3 nohz_full=2-3"
```

In particular, we instruct Linux to dedicate 16GB of RAM to 16 1GB-sized HugePages, to use SR-IOV pass-through (pt) mode, to disable C-states (low-power modes), and to isolate CPU cores 2 and 3 (i.e., no scheduling of interrupts or other processes on these two cores). We then manually pin Snort® threads to cores 2 and 3.

A typical command to run PktGen with a synthesized workload looks as follows:

```
sudo pktgen -l 0-1 -n 2 -m 4096 -- -P -m 1.0 -f test_64B_256F.lua
```

The command initializes PktGen to run GUI on core 0, one DPDK thread on core 1 (handling RX/TX queues on port 0), with 2 memory channels and 4GB of HugePages RAM. PktGen enables promiscuous mode on all ports and reads parameters of the workload from file `test_64B_256F.lua`. This file contains a Lua script to start sending 64B packets in 256 TCP flows on port 0.

A typical command to run PktGen with a PCAP workload looks very similar:

```
sudo pktgen -l 0-1 -n 2 -m 4096 -- -P -m 1.0 -f test_start.lua -s 0:smallFlows.pcap
```

Here we add `-s` to specify that we want to stream contents of a PCAP file on port 0. The script `test_start.lua` simply starts transmission.

A typical command to run Snort® or SEC-IDS:

```
sudo -E LD_LIBRARY_PATH="$LD_LIBRARY_PATH" snort® --daq dpdk -i dpdk0 --daq-var dpdk_args="-n 2 -l 1 -m 4096" –z 3 –c snort®.conf –R community_100.rules –A fast
```

The command executes Snort® with root privileges and passes `LD_LIBRARY_PATH` (required to find library dependencies). The arguments instruct Snort® to use `dpdk` DAQ module and call interface `dpdk0` (by convention). DPDK arguments are specified In `dpdk_args` and are similar to the ones above. We also specify three threads to run (one for DPDK, two for Snort®), configuration file `snort®.conf`, rules file `community_100.rules` with 100 rules, and the `fast` alert output.

## Appendix 4: Additional Experiments

For completeness, we report evaluation results for percentage of dropped and analyzed packets.

Figures 4.1 – 4.5 show the percentage of dropped packets as reported by DPDK. Vanilla Snort® almost never drops packets, i.e., the rate of network packets' consumption is higher than the capacity of the DPDK receive queue. Surprisingly, SEC-IDS drops 5-20% of packets in all configurations – except the "useless mode" without a configuration file at all – without any particular pattern or cause. Upon further investigation, it turned out that the startup phase of SEC-IDS (first 6-9 seconds of its execution) saw drop rates close to 100% while during the normal execution the drop rate never exceeded 1%. This behavior is understandable: during startup, SEC-IDS actively swaps pages in and out, with the Intel® SGX paging daemon `ksgxswapd` consuming 5-50% of CPU time. See Appendix 5 for example statistics.

Figures 4.6 – 4.10 show the percentage of analyzed packets as reported by Snort®. These figures exhibit the same patterns as our main evaluation (Figures 5—10). In general, by adding more Snort® threads, it would be possible to analyze 100% packets. We project that 8-10 Snort® threads and a single DPDK thread would be sufficient to saturate a 10Gbps link (conditioned on the larger EPC size available in next-generation Intel® SGX machines).

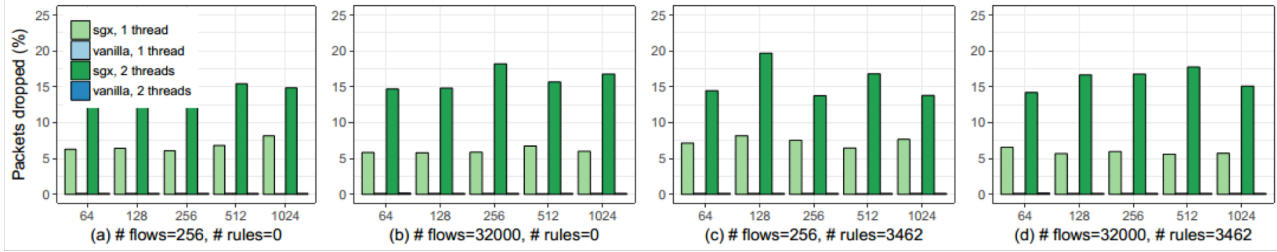

*Figure 4.1: Percentage of dropped packets of SEC-IDS and vanilla SGX with increasing packet size.*

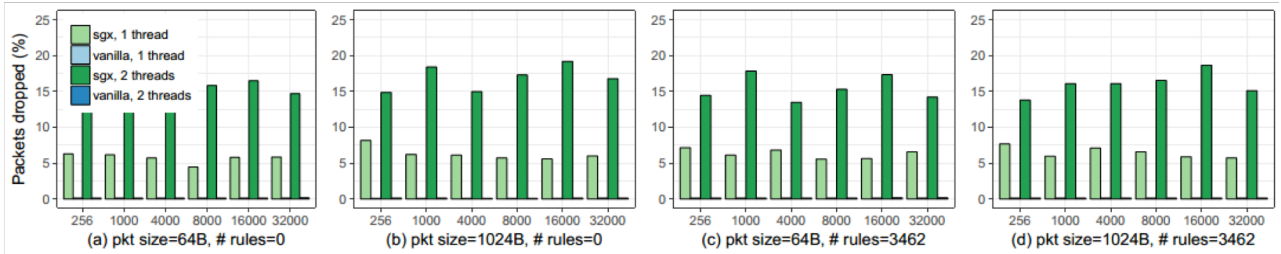

*Figure 4.2: Percentage of dropped packets of SEC-IDS and vanilla SGX with increasing number of TCP flows.*

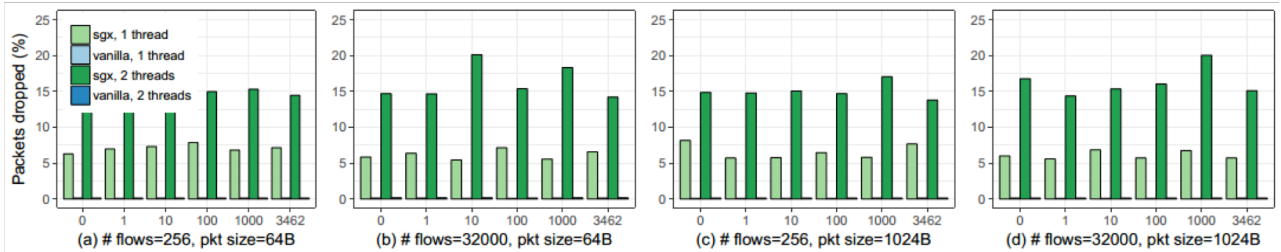

*Figure 4.3: Percentage of dropped packets of SEC-IDS and vanilla SGX with increasing number of rules (max 3462).*

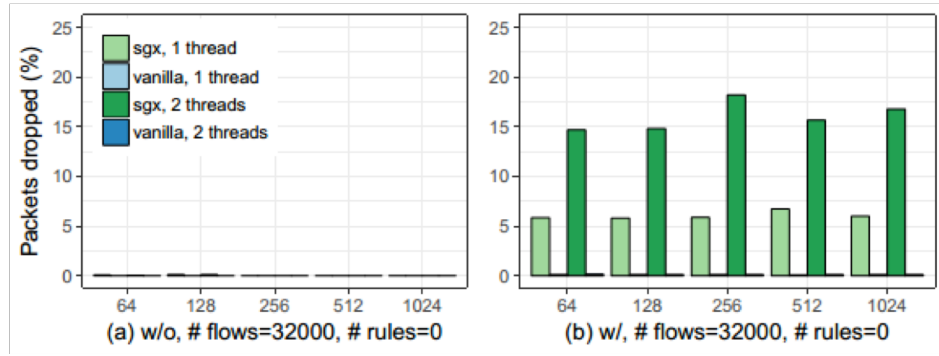

*Figure 4.4: Percentage of dropped packets of SEC-IDS and vanilla SGX without and with configuration file.*

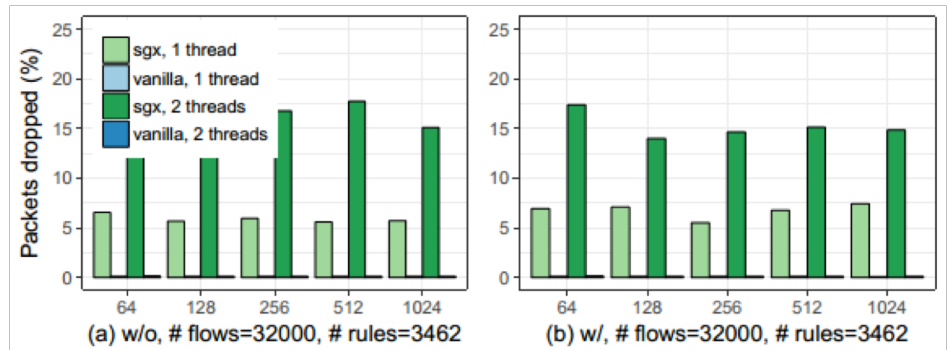

*Figure 4.5: Percentage of dropped packets of SEC-IDS and vanilla SGX without and with outputting alerts.*

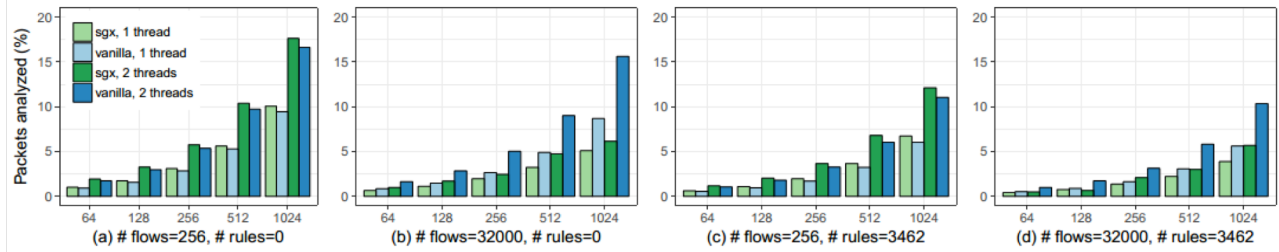

*Figure 4.6: Percentage of analyzed packets of SEC-IDS and vanilla SGX with increasing packet size.*

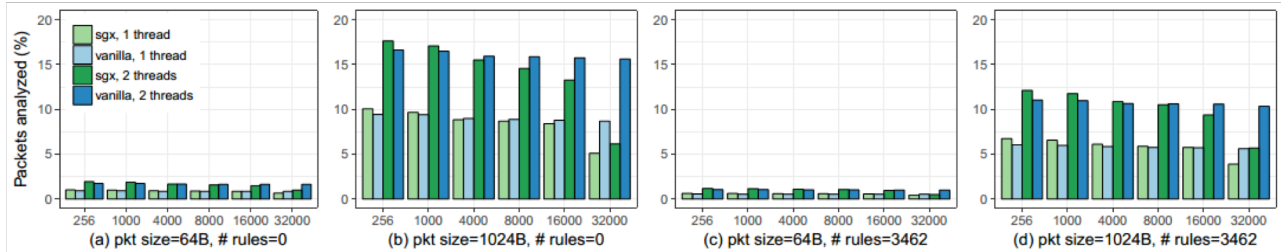

*Figure 4.7: Percentage of analyzed packets of SEC-IDS and vanilla SGX with increasing number of TCP flows.*

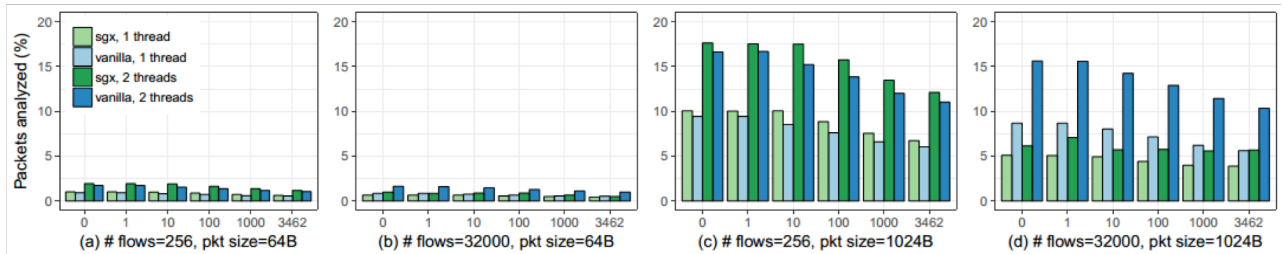

*Figure 4.8: Percentage of analyzed packets of SEC-IDS and vanilla SGX with increasing number of rules (max 3462).*

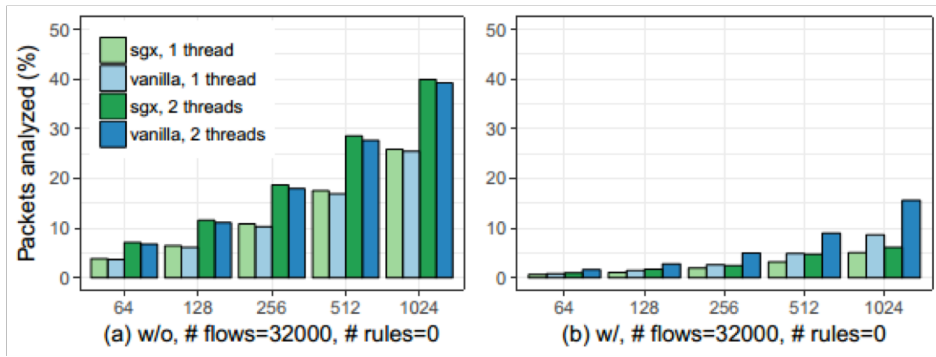

*Figure 4.9: Percentage of analyzed packets of SEC-IDS and vanilla SGX without and with configuration file.*

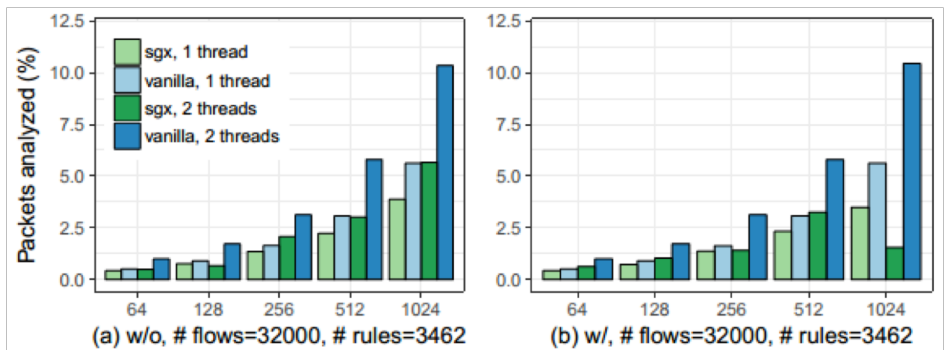

*Figure 4.10: Percentage of analyzed packets of SEC-IDS and vanilla SGX without and with outputting alerts.*

# Appendix 5: Example of Drop Rate and CPU Utilization

As shown in Appendix 4 and Figures 4.1-4.5, the drop rate of SEC-IDS is very high, ranging from 5% to 20%. Our suspicion was that the startup phase of SEC-IDS caused the skew in the number of dropped packets. The root cause is that at startup, the enclave needs to initialize a lot of code and data, which leads to a high rate of EPC paging. The following table shows one example run with a 3-second breakdown of (1) drop rate and (2) CPU utilization of EPC paging, i.e., `ksgxswapd`.

| Second | Drop rate, % | ksgxswapd CPU utilization, % |
| --- | --- | --- |
| 0 | 0.0 | 0.0 |
| 3 | **100.0** | **49.3** |
| 6 | **100.0** | **19.7** |
| 9 | 0.95 | 0.0 |
| 12 | 0.21 | **4.7** |
| 15 | 0.78 | **3.0** |
| 18 | 0.15 | **27.2** |
| 21 | 0.0 | 0.0 |
| 24 | 0.0 | 0.0 |
| 27 | 0.0 | 0.0 |
| 30 | 0.0 | 0.0 |
| 33 | 0.0 | 0.0 |
| 36 | 0.0 | 0.0 |
| 39 | 0.0 | 0.0 |
| 42 | 0.28 | 0.0 |
| 45 | 0.0 | 0.0 |
| 48 | 0.0 | 0.0 |
| 51 | 0.90 | 0.0 |
| 54 | 0.79 | 0.0 |
| 57 | 0.67 | 0.0 |
| 60 | 0.98 | 0.0 |

*Table 5.1: Example run for the SEC-IDS configuration: 2 threads, 1024B packet size, 256 flows, 3462 rules.*

To remove this skew from the drop rate Figures 4.1-4.5, we could re-design our experiments such that SEC-IDS is first started, warmed up for several seconds, and only then is loaded by PktGen. Fortunately, our current numbers on throughput (Figures 5-10) are still correct since these statistics were collected by Snort® after it went into steady state.